\RequirePackage{fix-cm}
\RequirePackage{fixltx2e}
\documentclass[10pt,a4paper,english,journal=jctcce,manuscript=article,layout=twocolumn]{achemso}
\usepackage[T1]{fontenc}
\usepackage[latin9]{inputenc}
\usepackage{color}
\usepackage{babel}
\usepackage{verbatim}
\usepackage{amsmath}
\usepackage{amssymb}
\usepackage{graphicx}

\DeclareMathSizes{10}{12}{12}{12}

\author{Alejandro Gil-Ley}
\author{Sandro Bottaro}
\author{Giovanni Bussi}
\email{bussi@sissa.it}
\affiliation[SISSA]{Scuola Internazionale Superiore di Studi Avanzati (SISSA), via Bonomea 265, 34136, Trieste, Italy}
\title[RNA]
{Empirical corrections to the Amber RNA force field with Target Metadynamics}

\usepackage{geometry}
\usepackage{breqn}
\usepackage{xkeyval} 
\usepackage{abstract}
\usepackage{changepage}
\usepackage{colortbl}
\usepackage{multirow}
\definecolor{lightgray}{gray}{0.9}
\usepackage{lineno}
\newcommand*{\sometext}{
The computational study of conformational transitions in nucleic acids still faces many challenges. 
For example, in the case of single stranded RNA tetranucleotides, agreement between simulations
and experiments is not satisfactory due to inaccuracies in the force fields commonly used in molecular
dynamics simulations. We here use experimental data collected from high-resolution X-ray structures to attempt an improvement of the latest version of the AMBER force field. A modified metadynamics algorithm is used to calculate correcting potentials designed to enforce experimental distributions of backbone torsion angles. Replica-exchange simulations of
tetranucleotides including these correcting potentials show significantly better agreement with
independent solution experiments for the oligonucleotides containing pyrimidine bases.
Although the proposed corrections do not seem to be portable to generic RNA systems,
the simulations revealed the importance of the $\alpha$ and $\zeta$ backbone angles on the modulation of the RNA conformational ensemble. The correction protocol presented here suggests a systematic procedure for force-field refinement.
}

\let\oldmaketitle\maketitle
\let\maketitle\relax

\makeatother

\begin{document}
\noindent \twocolumn[ 
 \begin{@twocolumnfalse} 
   \oldmaketitle 
   \begin{abstract} 
     \sometext 
     \\
   \end{abstract} 
 \end{@twocolumnfalse} 
]

\section*{Introduction}

Molecular dynamics is a powerful tool that can be used as a virtual
microscope to investigate the structure and dynamics of biomolecular
systems.\cite{dror2012biomolecular} However, the predictive power
of molecular dynamics is typically limited by the accuracy of the
employed energy functions, known as force fields. Whereas important
advances have been made for proteins,\cite{lindorff2010improved,lindorff2012systematic}
their accuracy for nucleic acids is still lagging behind.\cite{sponer2014molecular,bergonzo2015highly}
Force fields for RNA have been used since several years in many applications
to successfully model the dynamics around the experimental structures.\cite{cheatham2013twenty}
Traditionally, the functional form and parameters of these energy
functions have been assessed by checking the stability of the native
structure. This has lead for instance to the discovery of important
flaws in the parametrization of the backbone\cite{perez2007refinement}
and of the glycosidic torsion.\cite{zgarbova2011refinement} However,
to properly validate a force field it is necessary to ensure that
the entire ensemble is consistent with the available experimental
data. This can be done only using enhanced sampling techniques or
dedicated hardware. Recent tests\cite{condon2015stacking,bergonzo2015highly}
have shown that state-of-the-art force fields for RNA are still not
accurate enough to produce ensembles compatible with NMR data in solution
in the case of single stranded oligonucleotides. Similar issues have
been reported for DNA and RNA dinucleosides.\cite{vokacova2009structure,nganou2016disagreement}

Previous studies have shown that the distribution of structures sampled
from the protein data bank (PDB) may approximate the Boltzmann distribution
to a reasonable extent\cite{butterfoss2003boltzmann,mackerell2004extending,morozov2004close,lindorff2010improved}
and could even highlight features in the conformational landscape
that are not reproduced by state-of-the-art force fields.\cite{brereton2015native,bottaro2016}
This has been exploited in the parametrization of protein force fields.
For example, a significant improvement of the force fields of the
CHARMM family has been obtained by including empirical corrections
commonly known as CMAPs based on distributions from the PDB.\cite{mackerell2004improved,buck2006importance}

In this work, we apply these ideas to the RNA field and show how it
is possible to derive force-field corrections using an ensemble of
X-ray structures. At variance with the CMAP approach, we here correct
the force field using a self-consistent procedure where metadynamics
is used to enforce a given target distribution.\cite{white2015designing,marinelli2015ensemble}
Correcting potentials are obtained for multiple dihedral angles using
the metadynamics algorithm in a concurrent fashion. Since the target distributions are multimodal, we also use a recently developed enhanced
sampling technique, replica exchange with collective-variable tempering
(RECT),\cite{gil2015} to accelerate the convergence of the algorithm.
The correcting potentials are obtained by matching the torsion distributions
for a set of dinucleoside monophosphates. The resulting corrections
are then tested on tetranucleotides where standard force field parameters
are known to fail in reproducing NMR data.

\section*{Methods}

In this Section we briefly describe the target metadynamics approach
and discuss the details of the performed simulations.

\subsection*{Targeting Distributions with Metadynamics}

Metadynamics (MetaD) has been traditionally used to enforce an uniform
distribution for a properly chosen set of collective variables (CV)
that are expected to describe the slow dynamics of a system.\cite{laio2002escaping}
However, it has been recently shown  that the algorithm can be modified
so as to target a preassigned distribution which is not uniform.\cite{white2015designing,marinelli2015ensemble}
In this way a distribution taken from experiments, such as pulsed
electron paramagnetic resonance, or from an X-ray ensemble, can be
enforced to improve the agreement of simulations with empirical data.
We refer to the method as target metadynamics (T-MetaD), following
the name introduced in ref \cite{white2015designing}. For completeness,
we here briefly derive the equations. It is also important to notice
that the same goal could be achieved using a recently proposed variational
approach. \cite{valsson2014variational,shaffer2016enhanced}

In our implementation of T-MetaD a history dependent potential $V(s,t)$
acting on the collective variable $s$ at time $t$ is introduced
and evolved according to the following equation of motion

\begin{equation}
\dot{V}(s,t)=\omega e^{\beta(\tilde{F}(s(t))-\tilde{F}_{\max})}e^{-\beta(\frac{V_{\max}}{D})}e^{-\frac{(s-s(t))^{2}}{2\sigma^{2}}}\label{eq:1}
\end{equation}
Here $\beta=1/k_{B}T$, $k_{B}$ is the Boltzmann constant, $T$ the
temperature, $\omega$ is the initial deposition rate of the kernel
function which is here defined as a Gaussian with width $\sigma$,
$\tilde{F}$(s) is the free energy landscape associated to the target
distribution, $\tilde{F}_{\mbox{max}}$ indicates the maximum value
of the function $\tilde{F}$, and $D$ is a constant damping factor.
The target distribution is thus proportional to $e^{-\beta\tilde{F}(s)}$. We
define $\omega=\frac{Dk_{B}T}{\tau}$ where $\tau$ is the characteristic
time of bias deposition. The term $e^{\beta(\tilde{F}(s)-\tilde{F}_{\max})}$
adjusts the height of the bias potential, making Gaussians higher
at the target free-energy maximum and lower at its minimum. This forces
the system to spend more time on regions where the targeted free-energy
is lower. We notice that a similar argument has been used in the past
to derive the stationary distribution of both well-tempered metadynamics,
where Gaussian height depends on already deposited potential,\cite{barducci2008well}
and of adaptive-Gaussian metadynamics, where Gaussian shape and volume
is changed during the simulation.\cite{branduardi2012metadynamics} The
subtraction of $\tilde{F}_{\max}$ sets an intrinsic upper limit for
the height of each Gaussian, thus avoiding the addition of large forces
on the system. We notice that other authors used terms such as the
minimum of $F$ or the partition function to set an intrinsic lower
limit for the prefactor $e^{\beta(\tilde{F}(s)-\tilde{F}_{\max})}$.
\cite{white2015designing,marinelli2015ensemble}
At the same time, the term $e^{-\beta(\frac{V_{\max}}{D})}$ acts
as a global tempering factor\cite{dama2014well} and makes the Gaussian
height decrease with the simulation time so as to make the bias potential
converge instead of fluctuating. As observed in ref \cite{white2015designing},
the tempering approach used in well-tempered MetaD in this case would
lead to a final distribution that is a mixture of the target one with
the one from the original force field. For this reason, we prefer
to use here a global tempering approach.\cite{dama2014well} 

In the long time limit (quasi-stationary condition) the bias potential
will on average grow as\cite{barducci2008well,dama2014well} 

\begin{flushright}
\begin{multline}
\left\langle \dot{V}(s)\right\rangle =\intop ds'\omega e^{\beta(\tilde{F}(s')-\tilde{F}_{\max})}e^{-\beta(\frac{V_{\max}}{D})}\\
\begin{aligned}\end{aligned}
e^{-\frac{(s'-s)^{2}}{2\sigma^{2}}}P(s')
\end{multline}

\par\end{flushright}

\noindent where $P(s)$ is the probability distribution of the biased
ensemble. Defining the function $g(s')=\omega e^{\beta(\tilde{F}(s')-\tilde{F}_{\max})}e^{-\beta(\frac{V_{\max}}{D})}$
we can see this equation is a convolution of a Gaussian and a positive
definite function.
\begin{equation}
\left\langle \dot{V}(s)\right\rangle =\intop ds'e^{-\frac{(s'-s)^{2}}{2\sigma^{2}}}g(s')P(s')\label{eq:3}
\end{equation}

\noindent As shown in ref \cite{barducci2008well,dama2014well} this
average should be independent of $s$ in stationary conditions, so
that the function $g(s')P(s')$ should be also independent of $s'$,
though still dependent on time

\noindent 
\begin{equation}
\omega e^{\beta(\tilde{F}(s(t))-\tilde{F}_{\max})}e^{-\beta(\frac{V_{\max}}{D})}P(s)=C(t)
\end{equation}

\noindent By recognizing that $\tilde{F}_{\max}$ and $V_{\max}$
do not depend on $s$, one can transform the last equation to 

\noindent 
\begin{equation}
e^{\beta\tilde{F}(s)}P(s)=C'(t)
\end{equation}

\noindent which implies that 

\noindent 
\begin{equation}
P(s)\propto e^{-\beta\tilde{F}(s)}
\end{equation}

\noindent Thus, the system will sample a stationary distribution of
$s$ that  is identical to the enforced one.

Whereas the equations are here only described for a single CV, this
method can be straightforwardly applied to multiple CVs in a concurrent
manner. In this case, the total bias potential is the sum of the one-dimensional bias potentials 
applied to each degree of freedom. Indeed, similarly to the concurrent metadynamics used in RECT,\cite{gil2015}
all the distributions are self-consistently enforced.\cite{marinelli2015ensemble}
This is particularly important when biasing backbone torsion angles
in nucleic acids since they are highly correlated.\cite{saenger1984principles,richardson2008rna}
In this situation it is also convenient to use a biasing method that
converges to a stationary potential through a tempering approach,
to include in the self-consistent procedure of MetaD an additional
effective potential associated to the correlation between the dihedral
angles that is as close as possible to convergence.

\subsection*{Simulation Protocols}

\subsubsection*{RNA dinucleoside monophosphates}

Fragments of dinucleoside monophosphate with the sequence CC, AA,
CA, and AC were extracted from the PDB database of RNA X-ray structures
at medium and high resolution (resolution < 3 $\textrm{\AA}$). The
selected structures were protonated using \textit{pdb2gmx} tool from
GROMACS 4.6.7.\cite{hess2008gromacs} Free-energy profiles along the
backbone dihedral angles were calculated with the \textit{driver}
utility of PLUMED 2.1.\cite{tribello2014plumed}

Molecular dynamics simulations of the chosen RNA dinucleoside monophosphate
sequences were performed using the Amberff99bsc0\textit{$_{\chi OL3}$}
force field (named here Amber14).\cite{cornell1995second,perez2007refinement,zgarbova2011refinement}
The systems were solvated in an octahedron box of TIP3P water molecules
\cite{jorgensen1981quantum} with a distance between the solute and
the box wall of 1 nm. The system charge was neutralized by adding 1 Na$^{+}$ counterion.
The LINCS\cite{hess1997lincs} algorithm was used to constrain all
bonds containing hydrogens and equations of motion were integrated
with a timestep of 2 fs. All the systems were coupled to a thermostat
through the stochastic velocity rescaling algorithm.\cite{Bussi:2007aa}
For all non-bonded interactions the direct space cutoff was set to
0.8 nm and the electrostatic long-range interactions were treated
using the default particle-mesh Ewald\cite{darden1993particle} settings.
An initial equilibration in the NPT ensemble was done for 2 ns, using
the Parrinello-Rahman barostat.\cite{parrinello1981polymorphic} Production
simulations were ran in the NVT ensemble. All the simulations were
run using GROMACS 4.6.7\cite{hess2008gromacs} patched with a modified
version of the PLUMED 2.1 plugin.\cite{tribello2014plumed} 

T-MetaD simulations were run to enforce the probability distributions
of the angles $\epsilon_{1}$, $\zeta_{1}$, $\alpha_{2}$ and $\beta_{2}$
(see Fig. \ref{fig:1}), which were calculated from the X-ray fragments.
The target free-energy profiles were calculated with PLUMED 2.1. Distributions
were estimated as combination of Gaussian kernels, with a bandwidth
of 0.15 rad, and written on a grid with 200 bins spanning the $(-\pi,\pi)$
range. The bias potential used for the T-MetaD was grown using a characteristic
time $\tau=200$ ps and a dampfactor $D=100$. Gaussians with a width
of 0.15 rad were deposited every $N_{G}=$ 500 steps. 

We underline that simulations performed using T-MetaD could be non
ergodic for two reasons. First, there could be significant barriers
acting on CVs that are not targeted and thus not biased at this stage
(e.g. $\chi$ dihedral angles). Second, if the enforced distribution
of a CV is bimodal it will be necessary to help the system in exploring
both modes with the correct relative probability. It is thus necessary
to combine the T-MetaD approach with an independent enhanced-sampling
scheme. Here we used RECT, a replica exchange method where a group
of CVs is biased concurrently using a different bias factor for each
replica and one reference replica is used to accumulate statistics.\cite{gil2015}
When T-MetaD and RECT are combined, in each replica a T-MetaD is run
with the same settings, including the reference replica. The T-MetaD/RECT
simulation was run with 4 replicas for 1 $\mu$s each. For each residue
the dihedrals of the nucleic acid backbone ($\alpha$, $\beta$, $\gamma$,
$\epsilon$, $\zeta$), together with one of the Cartesian coordinates
of the ring puckering\cite{huang2014improvement} ($Zx$) and the
glycosidic torsion angle ($\chi$) were chosen as accelerated CVs
(see Fig. \ref{fig:1}). To help the free rotation of the nucleotide
heterocyclic base around the glycosidic bond, the distance between
the center of mass of nucleobases was also biased. For the dihedral
angles the Gaussian width was set to 0.25 rad and for the distance
it was set to 0.05 nm. The Gaussians were deposited every $N_{G}=$
500 steps. The initial Gaussian height was adjusted to the biasfactor
$\gamma$ of each replica, according to the relation $h=\frac{k_{B}T\left(\gamma-1\right)}{\tau_{B}}N_{G}\Delta t$,
in order to maintain the same $\tau_{B}=12$ ps across the entire
replica ladder. The biasfactor $\gamma$ ladder was chosen in the
range from 1 to 2, following a geometric distribution. In replicas
with $\gamma\neq1$ the target free energy was scaled by a factor
$1/\gamma$. Exchanges were attempted every 200 steps. Statistic was
collected from the unbiased replica. A sample input file is provided
as supplementary material (see Fig S1). 

Finally, a new RECT simulation was run for each dinucleoside with
the bias potentials obtained from the T-MetaD applied statically on
each replica. These calculations represent the results obtained with
a force field that includes the corrections from the PDB distributions
and are thus labeled as Amber$_{pdb}$. Statistics from these simulations
were collected to evaluate the effects of the corrections. The simulation
time was 1 $\mu$s per replica. 

\noindent \begin{center}
\begin{figure}
\noindent \begin{centering}
\includegraphics[scale=0.4]{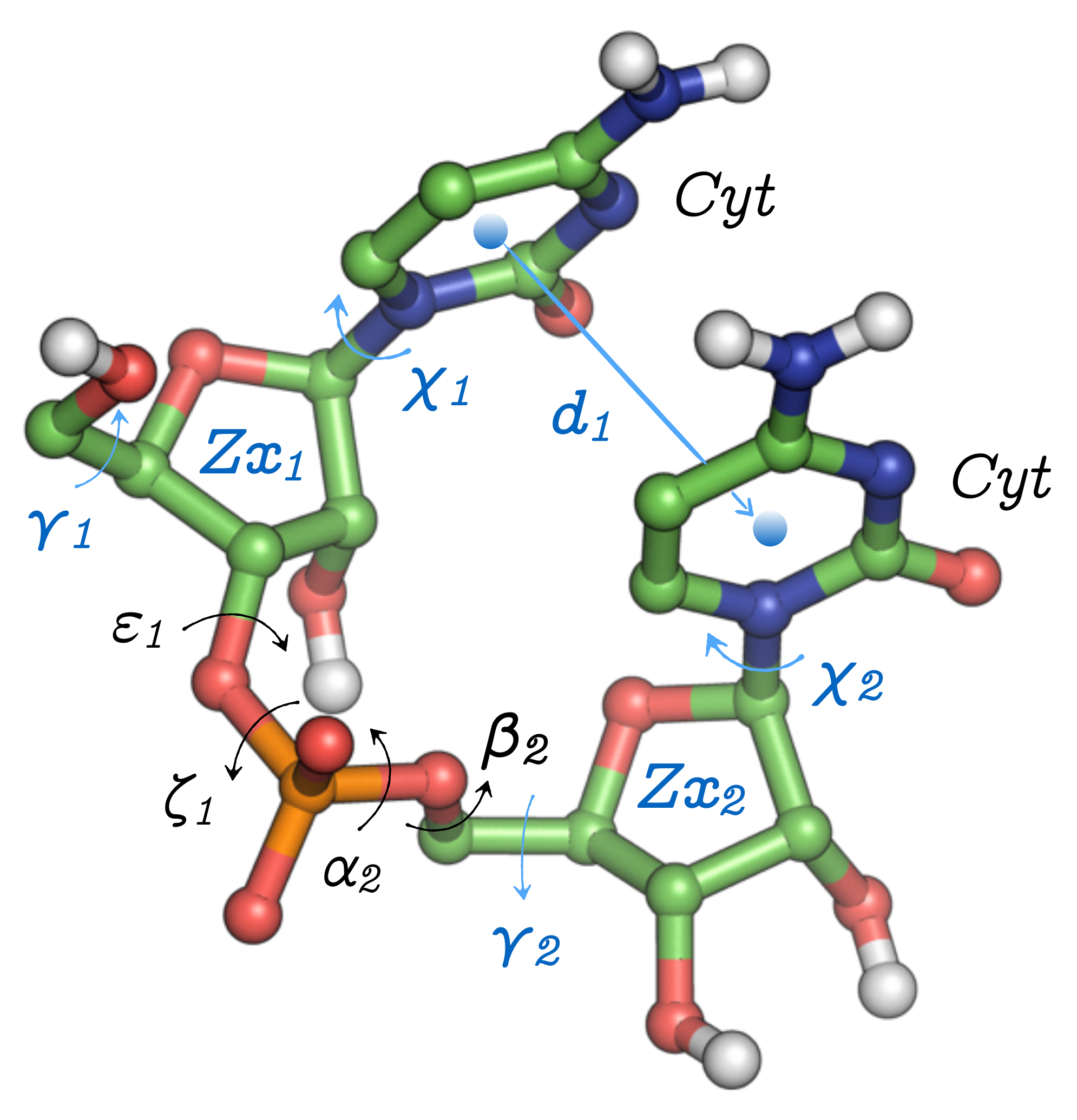}
\par\end{centering}

\noindent \centering{}\caption{\label{fig:1}Representation of a Cytosine-Cytosine dinucleoside monophosphate.
The backbone dihedrals selected for the force-field correction are
shown in black and the CVs accelerated in the RECT simulations are
shown in black or blue.}
\end{figure}

\par\end{center}

\subsubsection*{RNA Tetranucleotides}

To test the force field corrections derived on dinucleoside monophosphates,
temperature replica-exchange molecular dynamics (T-REMD) simulations
\cite{sugita1999replica} were performed on different tetranucleotide
systems with sequence CCCC, GACC and AAAA. The correcting potentials
calculated for the AA and CC dinucleosides were applied to all the
backbone angles of AAAA and CCCC tetranucleotides, respectively. For
the GACC tetranucleotide we combined the correcting potentials from
the T-MetaD simulations of AA, AC and CC, assuming a similarity between
purines A and G.

The T-REMD data related to the Amber14 force field and the protocol
for the new simulations performed using the Amber$_{pdb}$ force field
were taken from ref \cite{bottaro2016}. The systems were solvated
with TIP3P waters and neutral ionic conditions. We used 24 replicas
with a geometric distribution of temperatures from 300 to 400 K. Exchanges
were attempted every 200 steps. The simulation length was 2.2 $\mu$s
per replica.

\subsection*{Analysis}

The result of the molecular dynamics simulations was compared to NMR
experimental data of dinucleosides \cite{vokacova2009structure,olsthoorn1982influence,ezra1977conformational,lee1976conformational}
and tetranucleotides.\cite{yildirim2011benchmarking,tubbs2013nuclear,condon2015stacking}
$^{3}J$ vicinal coupling constants were calculated using Karplus
expressions.\cite{karplus1959contact,karplus1963vicinal} We took
into account the analysis made in refs \cite{sychrovsky2006calculation,vokacova2009,vokacova2009structure}
to select the most precise sets of parameters. Calculations were performed
using the software tool baRNAba \cite{bottaro2014role}. Details are
given in the supplementary information, subsection 1.1.

\section*{Results}

As a first step we used our approach to enforce the dihedral distribution
from the X-ray fragments on monophosphate dinucleosides AA, AC, CA,
and CC. Then, we show that the corrections are partly transferable
and could improve agreement with solution experiments for tetranucleotides.

\subsection*{Calculation of correcting potentials for dinucleoside monophosphates}

The Amber14 force field is considered to be one of the most accurate
ones for RNA, though it is failing to reproduce solution experiments
for short flexible oligomers. Recent benchmarks of different Amber
force field modifications based on reparametrization of the torsion
angles and non-bonded terms have shown that these changes did not
lead to a satisfactory agreement with solution experiments for tetranucleotides.\cite{bergonzo2015highly,condon2015stacking}
On the other hand, ensembles of tetranucleotides taken from the PDB
have a very good agreement with NMR data.\cite{bottaro2016} We thus
decided to add correcting potentials to the dihedral angle terms of
Amber14, based on information recovered from high-resolution X-ray
structures of RNA deposited in the PDB. We analyzed enhanced sampling
simulations of dinucleosides (described in this paper) and tetranucleotides
(described in a previous publication\cite{bottaro2016}), to select
a minimal amount of degrees of freedom to modify. This analysis indicated
the backbone angles $\epsilon$, $\zeta$, $\alpha$ and $\beta$
could benefit from a correction (a full description is presented in
supplementary information, section 2). We used T-MetaD to enforce
on those dihedrals the probability distributions obtained from fragments
of X-ray structures. RNA dinucleoside monophosphates were chosen as
model systems to obtain the correcting potentials. As the corrections
are sequence dependent, for each nucleobase combination we generated
an ensemble of experimental conformations from the PDB database that
had the same sequence as the dinucleoside monophosphates. 

\noindent \begin{center}
\begin{figure*}
\noindent \centering{}\includegraphics[scale=2]{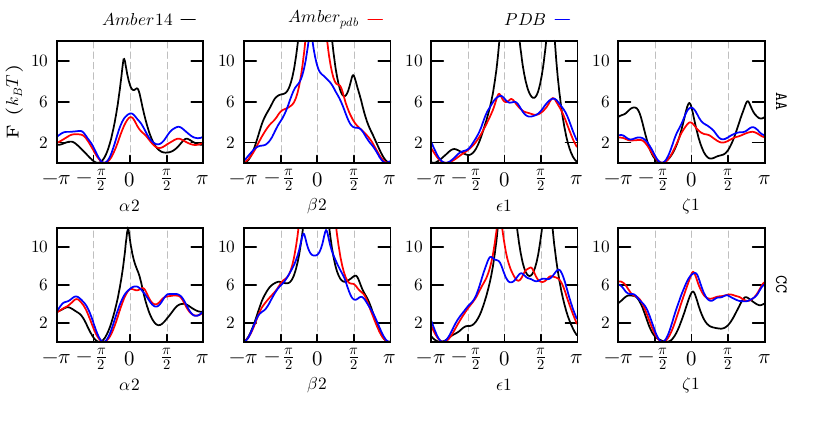}\caption{Free-energy profiles of backbone dihedral angles for the AA and CC
dinucleosides monophosphates from the X-ray ensemble (PDB) and the
RECT simulations with the standard force-field (Amber14) and the correcting
potential (Amber\textcolor{black}{\protect\textsubscript{pdb})}.\label{fig:2}}
\end{figure*}

\par\end{center}

In Fig. \ref{fig:2} we show the free energy profiles of AA and CC
dinucleosides projected on the $\epsilon$, $\zeta$, $\alpha$ and
$\beta$ angles. Amber14, Amber\textsubscript{pdb}, as well as the
target PDB ensembles are represented. The profiles of AC and CA are
shown in Fig S7. The similarity between the PDB and Amber\textsubscript{pdb}
profiles makes it clear that the corrections efficiently enforce the
distributions taken from the X-ray ensemble. Although some differences
are visible around the free-energy barriers, they are expected not
to be relevant for room temperature properties at equilibrium. Nevertheless,
the transition times and the behavior of the Amber\textsubscript{pdb}
potential at high temperatures could be affected by these barriers.
In general, barriers in the experimental ensemble are several $k_{b}T$
lower than those from the Amber14 force field. In the corrected ensemble
the multimodal character of the force field probability distributions
for the angles $\epsilon$, $\zeta$ and $\alpha$ is reduced, to
favor the conformations corresponding to the canonical A-form. The
observed agreement between the PDB and Amber\textsubscript{pdb} one-dimensional
probability distributions for the selected angles is not necessarily
translated into equivalence of the respective ensembles. This is seen
for example in the two-dimensional distributions shown in Figs S8-11.

Correcting potentials might in principle also affect the distribution
of non-biased degrees of freedom if the latter ones are correlated
with the former ones. The distribution of non-biased degrees of freedom,
such as the angles $\gamma$, $\chi$ and puckering coordinate $Z_{x}$,
is shown in Fig. S12. Overall, no difference is observed between
the Amber14 and Amber\textsubscript{pdb} free-energy profiles, with
the exception of the ratio between the C3'-\textit{endo} and C2'-\textit{endo}
conformations in CC. This is a consequence of the significant correlation
between the backbone angle $\epsilon$ and the puckering.

\begin{figure}[H]
\noindent \centering{}\includegraphics[scale=2]{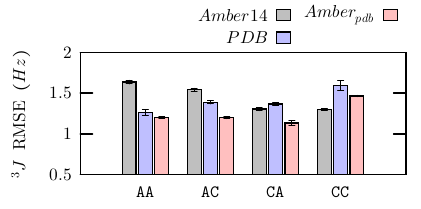}\caption{Agreement with the NMR $^{3}J$ vicinal coupling dataset of dinucleosides,
measured using the root mean square error (RMSE), for the ensembles
of X-ray structures (PDB), the Amber force field (Amber14) and the
corrected Amber force field (Amber\textcolor{black}{\protect\textsubscript{pdb}).}
Statistical errors were calculated using block averaging.\label{fig:3}}
\end{figure}

To asses the validity of the corrections, we compared all the ensembles
against NMR experimental data\cite{vokacova2009structure} (Fig \ref{fig:3}).
Individual $^{3}J$ vicinal coupling values from the experiments and
the simulations are reported in Table S2. In the case of AA, AC and
CA dinucleosides the agreement of Amber\textsubscript{pdb} with the
experimental data is better than that of Amber14 and of the X-ray
ensemble. This can be explained noticing that Amber\textsubscript{pdb}
combines the good agreement with NMR experiments of Amber14 for angles
in the nucleoside (dihedrals $\gamma,$ $\nu_{3}$ and $\chi$) with
that of the PDB distribution for angles in the backbone (dihedrals
$\epsilon$ and $\beta$), as shown in Fig S13. A notable exception
is the CC dinucleoside, where the correlation of backbone angles with
puckering mentioned above leads to slightly larger deviation in Amber\textsubscript{pdb}
with respect to Amber14. It should be noticed that the NMR observables
analyzed here cannot be used to directly determine the conformation
around the phosphodiester backbone ($\alpha$/$\zeta$), so the comparison
with the NMR $^{3}J$ vicinal coupling dataset does not take into account
the distribution of these angles. 

\noindent 

 We noticed that, whereas the NMR data was measured at 293 K (AA,
CA and AC) and 320 K (CC), simulations were performed at 300 K. However,
the agreement between the data for CC obtained at 320K and similar
NMR data obtained for a smaller number of couplings at 280K \cite{lee1976conformational}
shows that deviations induced by temperature changes are expected
to be much smaller than the typical deviations between molecular dynamics
and experiment observed here. It is also important to mention that
these RMSE values do not take into account systematic errors in the
Karplus formulas employed in this study. 

It is also interesting to measure the effect of the proposed backbone
corrections on the stacking interactions. Stacking free energies computed
according to the definition used in a recent paper\cite{condon2015stacking}
show that the correcting potential have barely no effect on stacking
(Fig S14). These numbers can also be compared with experimental values,\cite{lee1976conformational,ezra1977conformational,frechet1979thermalNN}
and indicate that Amber force field is likely overestimating stacking
interactions as suggested by several authors.\cite{chen2013high,brown2015stacking}
This comparison is however affected by the definition of stacked conformation,
which introduces a large arbitrariness in the estimation of stacking
free energies from MD.

\noindent \begin{center}
\begin{figure}
\noindent \centering{}\includegraphics[scale=2.5]{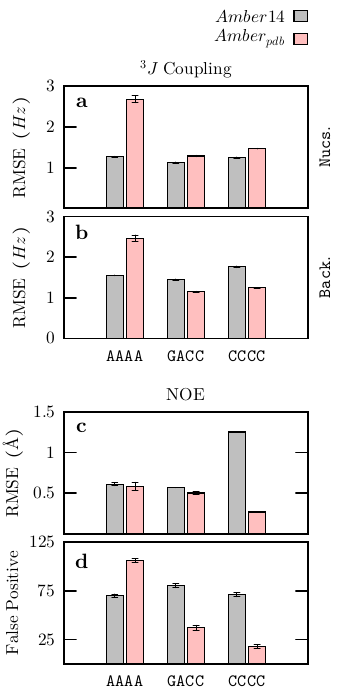}\caption{Agreement with the experimental $^{3}J$ vicinal couplings and NOE
distances of tetranucleotides. For the calculation of the $^{3}J$
RMSE the RNA torsion angles were divided in two groups: \textbf{a})
the dihedral angles in the ribose-ring region ($\chi$, $\nu$ and
$\gamma$) and \textbf{b}) the phosphate-backbone angles ($\epsilon$,
$\zeta$, $\alpha$ and $\beta$). In c) the RMSE between calculated
and predicted average NOE distances\textcolor{black}{{} is presented
and in }\textbf{\textcolor{black}{d}}\textcolor{black}{) it is shown
the number of false positives, i.e. the predicted distances below
5 $\protect\AA$ not observed in the experimental data.}\label{fig:4}}
\end{figure}

\par\end{center}

\subsection*{Validation of Amber\protect\textsubscript{pdb} potential on RNA
tetranucleotides}

\noindent \begin{center}
\begin{figure*}[!t]
\noindent \centering{}\includegraphics[scale=0.15]{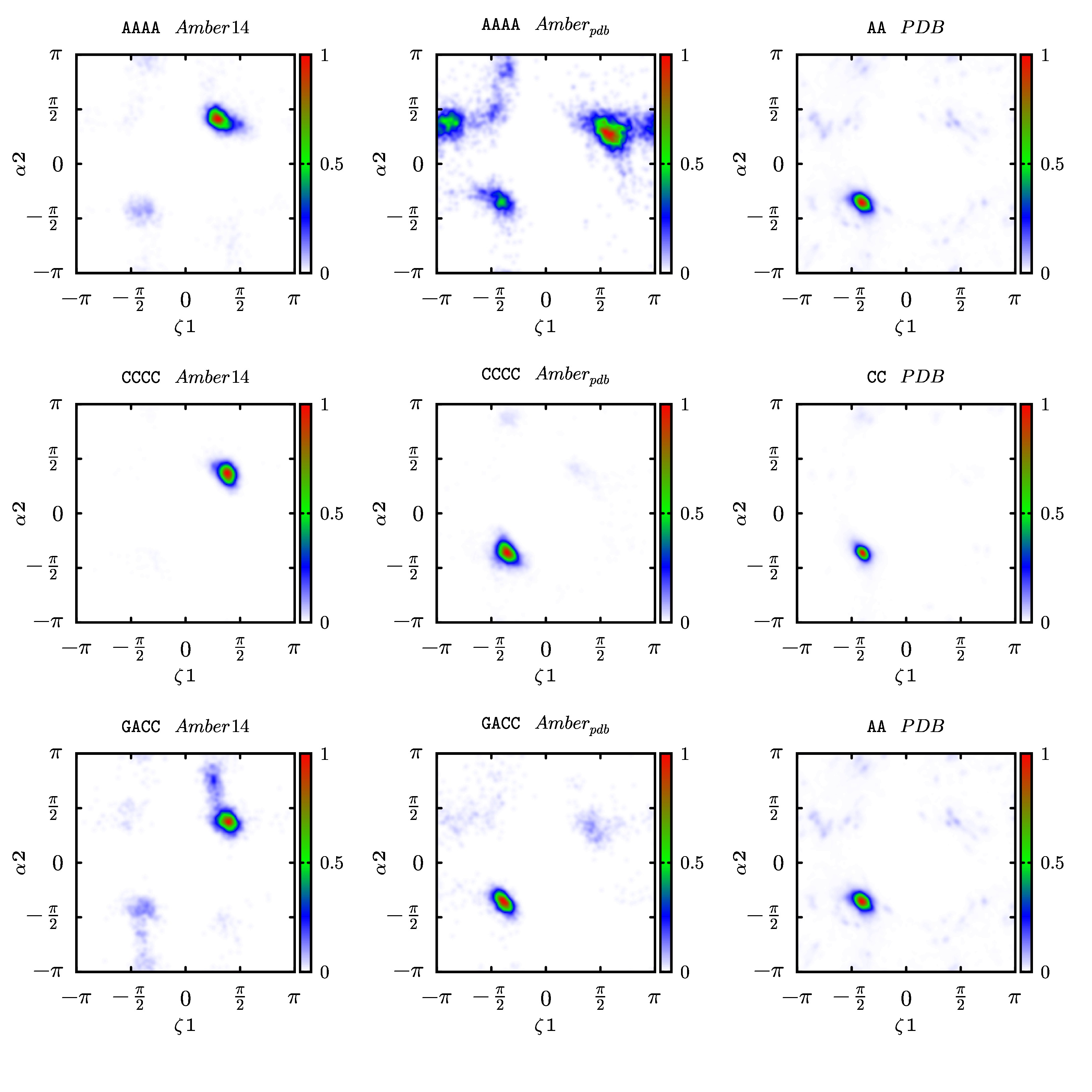}\caption{Probability distributions of the backbone dihedral angles of AAAA
and CCCC tetranucleotides, in the region between residue 1 and 2.
Results from the RECT simulations with the standard force-field (Amber14),
the correcting potential (Amber\textcolor{black}{\protect\textsubscript{pdb})
and the} dinucleoside X-ray ensembles (PDB)\textcolor{black}{{} used
to generate the correcting potentials.}\label{fig:5}}
\end{figure*}

\par\end{center}

The correcting potentials discussed above are designed so as to enforce
the PDB distribution on dinucleosides monophosphates. We here used
these corrections to perform simulations on larger oligonucleotides.
In particular, we performed extensive simulations of tetranucleotides,
which are considered as good benchmarks for force-field testing, as
their small size makes the generation of converged ensembles accessible
to modern enhanced sampling techniques. We performed three T-REMD
simulations with the Amber\textsubscript{pdb} potential for the tetranucleotide
sequences AAAA, GACC and CCCC. These systems have been used before
in very long (hundred of $\mu$s) simulations\cite{bergonzo2013multidimensional,henriksen2013reliable,roe2014evaluation,bergonzo2015highly,bergonzo2015improved}
and NMR experimental data is available.\cite{yildirim2011benchmarking,tubbs2013nuclear,condon2015stacking}
The Amber14 T-REMD data were taken from ref \cite{bottaro2016}. 

The $^{3}J$ coupling RMSE, the NOE-distance RMSE, and the number
of distance false positives, i.e. the MD predicted NOEs not observed
in the experiment, are presented in Fig \ref{fig:4}. For these systems
the number of false positives is one of the most important parameters
to assess the quality of the MD ensembles.\cite{condon2015stacking}
In the case of tetranucleotides containing pyrimidines (GACC and CCCC),
the correcting potential improves significantly the agreement with
the experimental data, mostly for the NOEs (see Fig S15). This is
confirmed by the root-mean-square deviation (RMSD) distribution shown
in Figure S16 where it can be appreciated that for these two sequences
the corrections lead to an overall improvement of the ensemble by
disfavoring the intercalated and inverted structures with a large
 RMSD from native. A completely different scenario is found for
the Amber\textsubscript{pdb} ensemble of AAAA, where the corrections
surprisingly diminish the agreement with experiments. This can be
also appreciated in a shift of the Amber\textsubscript{pdb} RMSD
distribution peaks to higher RMSD values due to an increase in the population
of compact structures (Fig S16). It should be noticed that the effect
of the correcting potentials in purines and pyrimidines depends strongly
on the sequence length. Whereas the AAAA tetranucleotide is negatively
affected by the corrections, the AA dinucleoside is the one that benefits
the most from them.

\noindent \begin{center}
\begin{figure}
\noindent \centering{}\hspace*{-0.4in}\includegraphics[scale=2.3]{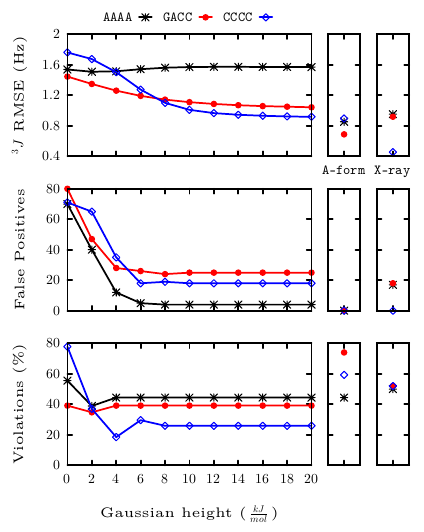}\caption{Agreement with the experimental data for the Amber14 reweighted ensemble
as a function of the Gaussian potential height. The bias potential
was centered on $\alpha$(\textit{g}+)/$\zeta$(\textit{g}+) conformation
($\frac{\pi}{2},\frac{\pi}{2}$) with a sigma per angle of 0.7 rad.
``A-form'' represent a canonical A-form structure and ``X-ray''
an ensemble of tetranucleotide fragments, with the same sequence,
from the PDB (all taken from ref \cite{bottaro2016}).\label{fig:6}}
\end{figure}

\par\end{center}

As discussed in the section 2 of the SI, the conformation along the
phosphodiester backbone is very different between compact and extended
tetranucleotide structures. The probability distribution maps of the
$\alpha_{2}$/$\zeta_{1}$ backbone dihedral angles from the tetranucleotides
T-REMD simulations and the dinucleosides X-ray ensembles used to generate
the corrections are depicted in Fig \ref{fig:5}. Only phosphodiester
backbone torsion angles are shown, because they are the ones mostly
affected by the correction. The other backbone angles maps are shown
in the SI (Figs S17-25). In the PDB ensembles the distributions are
always unimodal, independently of the sequence, with a peak at the
$\alpha$(\textit{g}-)/$\zeta$(\textit{g}-) conformation, whereas
in the Amber14 ensemble the $\alpha$(\textit{g}+)/$\zeta$(\textit{g}+)
and $\alpha$(\textit{g}-)/$\zeta$(\textit{g}-) conformations are
both significantly populated. The effects of the corrections, as seen
before, are highly sequence dependent. In case of GACC and CCCC, the
$\alpha$(\textit{g}-)/$\zeta$(\textit{g}-) rotamer is stabilized
in the Amber\textsubscript{pdb} distributions, with the population
of $\alpha$(\textit{g}+)/$\zeta$(\textit{g}+) significantly decreased
with respect to Amber14. On the contrary, for AAAA the $\alpha$(\textit{g}+)/$\zeta$(\textit{g}+)
conformation is not unfavored by the correcting potentials, despite
not being significantly present in the PDB ensemble. This could be
due to the fact that the one dimensional target free-energy profile
for dihedrals $\alpha$ and $\zeta$ for the AA (Fig \ref{fig:2})
exhibits barriers which are approximately 4 $k_{b}T$ smaller with
respect to the ones from the Amber14 force field. The effect of the
decreased barrier height can be appreciated in the $\alpha_{2}$/$\zeta_{1}$
probability distribution of AAAA, where the amount of torsional space
explored is increased by the corrections.

\subsection*{Consequences on future force field refinements}

The good agreement of the Amber\textsubscript{pdb} ensembles with
the NMR observables, in the case of CCCC and GACC tetranucleotides,
suggests that the RNA conformational space sampled by state-of-the-art
force field could be modified to better match experimental solution
data by penalizing rotamers of the $\alpha$ and $\zeta$ angles. As
a further test, we reweighted the T-REMD Amber14 ensembles with an
additional two-dimensional penalizing Gaussian potential centered
on the $\alpha$(\textit{g}+)/$\zeta$(\textit{g}+) conformation.
Results are shown in Fig \ref{fig:6} for different Gaussian heights.
Overall, the agreement with the NMR experimental data improves considerably
with respect to the original force field as the Gaussian height increases.
The relative population of the $\alpha$/$\zeta$ conformations has
an important impact on the number of false positive NOE contacts which
indicates the presence of intercalated structures. This improvement
is achieved without changing the non bonded interactions as it has
also been proposed.\cite{chen2013high} It is however important to
observe that these results are obtained by performing a reweighting,
and that corrections should be validated by performing separate simulations
with this bias potential.

\section*{Discussion}

In this paper we apply targeted metadynamics to sample preassigned
distributions taken from experimental data.\cite{white2015designing,marinelli2015ensemble}
At variance with the original applications, we here combine T-MetaD
with enhanced sampling showing that these protocols can also be used
when the investigated ensembles have non-trivial energy landscapes
separated by significant barriers .

We apply the method to RNA oligonucleotides, for which the Amber14
force field was proven to be in significant disagreement with solution
NMR data.\cite{yildirim2011benchmarking,henriksen2013reliable,tubbs2013nuclear,bergonzo2013multidimensional,condon2014optimization,condon2015stacking,bergonzo2015highly,bergonzo2015improved}
Since tetranucleotide fragments extracted from high resolution structures
in the PDB were shown to match NMR experiments better than Amber14
force field,\cite{bottaro2016} we here used X-ray structures to build
reference distributions of backbone dihedral angles that are then
used to devise correcting potentials. More precisely, we use T-MetaD
to enforce the empirical distribution of the dihedral angles in the
phosphate backbone ($\epsilon$, $\alpha$, $\zeta$ and $\beta$)
on four dinucleoside monophosphates.

We calculated the correcting potentials concurrently for all the four
angles in order to change the distribution of these consecutive dihedrals
along the backbone chain taking into account their correlation. The
method successfully enforced the distributions taken from the PDB
on all the angles. The new ensemble generated by the corrected force
field (Amber\textsubscript{pdb}) was independently validated against
solution NMR data that was not used in the fitting of the corrections.
For three of the four dinucleosides studied, Amber\textsubscript{pdb}
showed a better agreement with the NMR data compared with Amber14
and with the X-ray ensemble.

We then tested the portability of the correcting potentials by simulating
three tetranucleotides, GACC, CCCC and AAAA. In the case of GACC and
CCCC the agreement with NMR data is significantly improved by the
corrections. Surprisingly, for AAAA the corrections have the opposite
effect and increase the probability of visiting compact structures
making the simulated ensemble less compatible with solution experiments.
It should be noticed here that this is a non obvious result since
the PDB database is expected to have an intrinsic bias towards A-form
structures and should thus in principle increase the agreement with
solution experiments in this specific case. This indicates that porting
the corrections from dinucleosides to tetranucleotides is not straightforward
because the coupling between the multiple corrected dihedrals could
affect the resulting ensemble in an non-trivial way. Additionally,
corrections applied to dihedral angles alone might be not sufficient
to compensate errors arising from inexact parametrization of van der
Waals or electrostatic interactions.\cite{chen2013high} Overall,
the tests we performed indicate that the corrections derived here
should not be considered as portable corrections for the simulation
of generic RNA sequences.

Nevertheless, by comparing the backbone angle distributions on the
different RNA simulations and the X-ray ensembles, we were able to
find possible hints pointing at where refinement of dihedral potentials
could lead to an advancement in RNA force fields. In this respect,
the results for GACC an CCCC show the significant improvement observed
in the Amber\textsubscript{pdb} simulations for those systems could
be reproduced by simply penalizing the $\alpha$(\textit{g}+)/$\zeta$(\textit{g}+)
conformation, which is overpopulated in Amber14. By a straightforward
reweighting procedure, we showed that simple Gaussian potentials that
disfavor this conformation significantly improved the experimental
agreement with solution experiments for all the three tetranucleotides.
 Recent modifications of the Lennard-Jones parameters for phosphate
oxygens\cite{steinbrecher2012revised} and different water models\cite{bergonzo2015improved}
were shown to affect the conformational ensemble of RNA tetranucleotides.\cite{bergonzo2015highly,bergonzo2015improved}
It might be interesting to combine these modified parameters for non-bonded
interactions with the here introduced procedure for dihedral angle
refinement. 

The nature of the correction methodology discussed in this paper is
very different from the classical approach to force field parametrization,
as it aims to correct the free energy of the system, instead of fitting
the potential energy landscape of the dihedral angles while constraining
the other degrees of freedom.
It is important to notice that the dihedral angle distributions taken from the fragments of the PDB structures do not necessarily
represent the conformational ensembles of dinucleosides or tetranucleotides
in solution.
Indeed, some of the interaction patterns that are present in large structures
crystallized in the PDB do not exist in short oligonucleotides.
For this reason, in this work the distributions were validated against independent solutions NMR experiments.
This allowed the
dihedral angles from the PDB distributions that performed better than the force field to be identified.
We also recall that in our procedure
the force-field torsion energy function is not refitted, but a bias potential is added to the total energy of the system in order to 
match the free-energy profile of the torsion angles with target ones.
Thus, a major advantage of this approach
is that it takes explicitly into account the entropic contributions, the cross correlations between torsional angles, and inaccuracies in the non-bonded interactions, among other effects.

\section*{Conclusion}

In conclusion, in this work we applied the target metadynamics protocol
to modify dihedral distributions in dinucleosides. The procedure successfully
enforces reference distributions taken from the PDB without affecting
the distribution of the dihedral angles that were not biased. However,
the attempt to port these corrections to tetranucleotides lead to
ambiguous results when applied to different sequences. This could
be partly due to the fact that distribution form the PDB are not necessarily
a good reference for refinement.

Nevertheless, the simulations revealed the importance of the $\alpha$/$\zeta$
angles rotamers on the modulation of the conformational ensemble,
and that by only penalizing the $\alpha$(\textit{g}+)/$\zeta$(\textit{g}+)
rotamer the quality of the ensemble is significantly improved to levels
not reported before. 

\acknowledgement

Thomas Cheatham III, Fabrizio Marinelli, and Ji{\v r}{\'\i} {\v S}poner are acknowledged for carefully
reading the manuscript and providing several useful suggestions. The
research leading to these results has received funding from the European
Research Council under the European Union\textquoteright s Seventh
Framework Programme (FP/2007-2013) / ERC Grant Agreement n. 306662,
S-RNA-S.

\bigskip{}

\noindent 

\providecommand{\latin}[1]{#1}
\providecommand*\mcitethebibliography{\thebibliography}
\csname @ifundefined\endcsname{endmcitethebibliography}
  {\let\endmcitethebibliography\endthebibliography}{}

\end{document}